\documentclass[twocolumn,aps,pra,showpacs,superscriptaddress,tightenlines]{revtex4-1}
\usepackage{amsmath}
\usepackage{mathrsfs}
\usepackage{amsfonts}
\usepackage{graphicx}
\usepackage{epsfig}
\usepackage{color}
\usepackage[colorlinks,citecolor=blue]{hyperref}

\begin{document}
	
\title{Nonreciprocal photon bundle emission}

\author{Baijun Li}
\affiliation{Research Center for Quantum Physics, Huzhou University, Huzhou 313000, China}

\affiliation{Institute of Quantum Precision Measurement, State Key Laboratory of Radio Frequency Heterogeneous Integration, College of Physics and Optoelectronic Engineering, Shenzhen University, Shenzhen 518060, China}

\affiliation{Quantum Science Center of Guangdong-Hong Kong-Macao Greater Bay Area (Guangdong), Shenzhen 518045, China}

\author{Jing-Xue Liu}
\affiliation{School of Physics, Henan Normal University, Xinxiang 453007, China}

\author{Tian-Xiang Lu}
\affiliation{College of Physics and Electronic Information, Gannan Normal University, Ganzhou 341000, China}

\author{Le-Man Kuang}
\affiliation{Key Laboratory of Low-Dimensional Quantum Structures
and Quantum Control of Ministry of Education, Department of
Physics and Synergetic Innovation Center for Quantum Effects and
Applications, Hunan Normal University, Changsha 410081, China}

\author{Chaohong Lee}
\affiliation{Institute of Quantum Precision Measurement, State Key Laboratory of Radio Frequency Heterogeneous Integration, College of Physics and Optoelectronic Engineering, Shenzhen University, Shenzhen 518060, China}

\affiliation{Quantum Science Center of Guangdong-Hong Kong-Macao Greater Bay Area (Guangdong), Shenzhen 518045, China}

\author{Hui Jing}\email{jinghui73@foxmail.com}
\affiliation{Key Laboratory of Low-Dimensional Quantum Structures
and Quantum Control of Ministry of Education, Department of
Physics and Synergetic Innovation Center for Quantum Effects and
Applications, Hunan Normal University, Changsha 410081, China}

\affiliation{College of Science, National University of Defense Technology, Changsha, 410073, China}

\begin{abstract}
Quantum squeezing, a cornerstone of quantum optics and photonics, has played a key role in achieving ultra-precision sensing and realizing nonreciprocal engineering.  
However, the nonreciprocal multiquanta emission has remained largely unexplored by using directional quantum squeezing. 
Here, the one-way photon-photon bundle emission in a compound system consisted of two coupled optical resonators and a two-level atom is investigated. 
It is found that the directional quantum squeezing induces the asymmetric frequency detuning and photon hopping interaction between the two resonators, leading to the directional excitation of the two-photon super-Rabi oscillation.
In particular, by harnessing intrinsic dissipation of the system, two types of two-photon bundle emission can be selectively induced for the probe field input from one direction while it is prohibited with the probe from the other direction. 
This finding bridges the broad fields ranging from nonreciprocal physics to quantum squeezing optics and multiquanta emission control through an all-optical approach, which can enable potential applications in chiral quantum emitters and backscattering-immune photonic communications.
\end{abstract}
		
\maketitle

\section{Introduction}

As a critical resource for quantum information processing~\cite{DellAnno06Multiphoton,Pan12Multiphoton}, multiphoton sources are indispensable in applications ranging from quantum lithography~\cite{DAngelo01Two,Boto00Quantum}, quantum communication~\cite{Kimble08The,Gisin07Quantum}, to quantum metrology~\cite{Giovannetti04Quantum,Giovannetti11Advances}. 
An important strategy for generating such sources employs the multiphoton bundle emission mechanism. 
In this mechanism, the fundamental unit of emission shifts from  single photon to a group of $N$ photons~\cite{Munoz14Emitters}. 
By engineering the dissipative channels, bundle emission enables the realization of both multiphoton lasers and photon guns. 
In recent years, multiphoton bundle emission has been predicted in various systems, such as cavity quantum electrodynamics (QED) systems~\cite{Munoz14Emitters,Munoz15Enhanced,Chang16Deterministic,Munoz18Filtering,Bin21Parity,Jiang23Multiple,Tang23Strong,Gou24Antibunched}, waveguide-QED systems~\cite{Tudela15Deterministic,Tudela17Efficient,Xing24Deterministic}, cavity optomechanical systems~\cite{Zou22Dynamical}, and superconducting circuit~\cite{Menard22Emission}. 
Notably, the emission of photon multiplets was experimentally demonstrated in a dc-Biased superconducting circuit~\cite{Menard22Emission} and emission of photon pair was shown by using a three-level ladder atom and two optical cavities~\cite{Chiarella24Two}. 
Furthermore, the framework of bundle emission has also been extended into various fields, encompassing phenomena including N-phonon and magnon bundle emission~\cite{Deng21Motional,Bin20Phonon,Yuan23Magnon,Bin24Entangled,Gou24Hybrid}. 
Specially, the combination of photon pairs and nonreciprocal physics~\cite{Sounas17Non,Peng14Parity,Shi15Limitations,Shen16Experimental,Maayani18Flying,Zhang18Thermal,Xia18Cavity,Cotrufo24Passive, Zhang25Observation,Li25Nonreciprocal}, termed nonreciprocal bundle emissions of quantum entangled pairs~\cite{Bin24Nonreciprocal}, has recently been implemented in a spinning cavity QED system, enabling the directional manipulation of multiquantum emitters. 

Quantum squeezed light is a potent form of nonclassical light and is considered a cornerstone of quantum optics~\cite{Walls83Squeezed,Slusher85Observation,Wu86Generation,Schnabel17Squeezed,Ulanov25Quadrature}. 
This kind of light is an essential resource for quantum metrology, as it allows measurement precision to exceed the shot-noise limit~\cite{Schnabel17Squeezed,Caves80On,Caves81Quantum}. 
In the advanced LIGO detectors, for instance, squeezed states are deployed to improve the sensitivity of the interferometers~\cite{Grote13First,Aasi13Enhanced,Tse19Quantum}.
Also, quantum squeezing plays a critical role in quantum information science~\cite{Braunstein05Quantum}. 
By dramatically enhancing light-matter interactions~\cite{Lu15Squeezed,Leroux18Enhancing,Qin18Exponentially,Qin24Exponentially}, it allows for the engineering of nonclassical states and the exponential improvement of the signal-to-noise ratio of the dispersive qubit readout.
In particular, quantum squeezing also demonstrates significant potential in nonreciprocal physics~\cite{Fruchart21Nonreciprocal,Huang18Nonreciprocal,Li19Nonreciprocal,Li21Nonreciprocal,Li24Loss,Ahmadi24Nonreciprocal,Veenstra24Non,Nadolny25Nonreciprocal,Zhu24Nonreciprocal,Zhang25Chirality}. 
Very recently, an integrable all-optical nonreciprocal transistor has been proposed, utilizing unidirectional quantum  squeezing~\cite{Tang22Quantum,Huang24Nonreciprocal}.
Moreover, the squeezing-based approach also has been extend widely to various one-way phenomena, including the nonreciprocal photon blockade~\cite{Shen23Tunable,Wang23Squeezing,Liu23Parametric}, one-way phonon/magnon laser~\cite{Huang22Parametric,Lu24Quantum}, and directional entanglement~\cite{Jiao20Nonreciprocal,Lu24Directional}. 
The connection between quantum squeezing and nonreciprocal multiquanta emission, however, is yet to be explored.

Here, we explore the potential of the directional quantum squeezing for generating the nonreciprocal two-photon bundle emission. 
Our proposed scheme involves a compound system comprising two coupled cavities and a two-level atom. 
One of the cavities is embedded with $\chi^{(2)}$ nonlinearity and driven by a strong coherent laser field.  
The strong pump field generates a chiral quantum squeezing, which in turn induces asymmetric frequency detuning and photon hopping interactions between the two resonators.  
Consequently, nonreciprocal two-photon super-Rabi oscillations emerge from the system dynamics. 
Furthermore, by leveraging the inherent system dissipation, such one-way super-Rabi oscillations enable the realization of two distinct types of directional two-photon bundle emission. 
Therefore, by combining the advantages of nonreciprocity and quantum bundle emission, nonreciprocal photon bundle emission enables more precise control over entangled photon pairs, ensuring that they are generated in specific directions and under controlled conditions. 
This work expands the exploration of all-optical nonreciprocal approach based on quantum squeezing into the multiquanta physics regime, holding potential applications in on-chip chiral quantum light sources and backscattering-immune integrated quantum communication components.

\section{Model and Nonreciprocal super-Rabi oscillation} \label{M and S}
\begin{figure*}[t!]
	\includegraphics[width=0.95\linewidth]{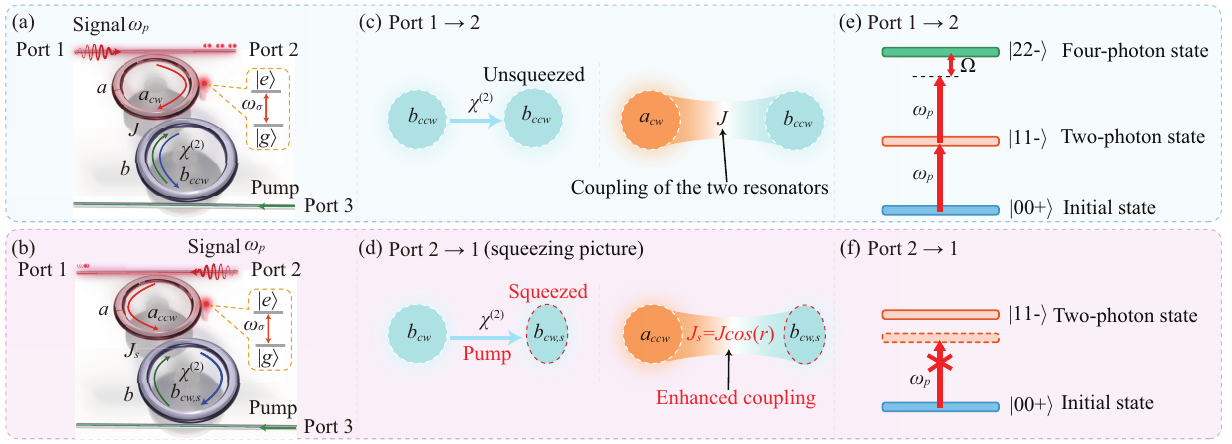}
	\centering
	\caption{Schematic of nonreciprocal two-photon bundle emission via directional quantum squeezing. 
		(a,b) Compound system comprises two coupled whispering-gallery mode microcavities and a two-level atom, where a weak probe field is input from port 1 or from port 2. 
		(c,d) Cavity $b$, containing $\chi^{(2)}$ nonlinearity, is driven by a strong coherent laser from port 3, generating the directed quantum squeezing. 
		A strong pump field also is applied to the two-level atom. 
		The directed squeezing leads to an asymmetric
		detuning and photon hopping interaction in the system.  
		The physical origin of the nonreciprocal two-photon bundle emission. 
		(e) When the probe field is input from the left side, the super-Rabi oscillation occurs because the frequencies match the resonances of the transition between $|00+\rangle$ and $|11-\rangle$. 
		Triggered by dissipation, photons can be emitted, leading to the emergence of photon bundle emission.
		(f) When the probe field is input from the right side, quantum squeezing modifies the effective coupling between the two resonators. 
		Thus, there is a frequency detuning between probe field and the resonance transmission and the photon bundle emission disappears.  
		Experimentally accessible parameters are chosen as $\Omega=\Delta_\sigma^2 + 4\Omega_\sigma^2$, $\Omega_\sigma = 6\,\xi$, $\Delta_{\sigma a}=\omega_\sigma - \omega_a = -23\,\xi, \Delta_{b a}=\omega_b - \omega_a = 2\,\xi, \Delta_b^d=\xi$, and $J=0.4\,\xi$.}
	\label{Fig1}
\end{figure*}

The scheme of nonreciprocal photon bundle emission is depicted in Figs.~\ref{Fig1} (a,b). 
We consider a compound system composed of two coupled  whispering-gallery-mode (WGM) microcavities (cavity $a$ and $b$) and two nearby optical fiber. 
The coupled WGM resonators system is experimentally feasible, which has led to various intriguing phenomena, such as ultra-low threshold phonon laser~\cite{Zhang18A} and nonreciprocal transmission~\cite{Peng14Parity}. 
The two optical modes within the resonators can interact via evanescent fields, and the coupling strength can be adjusted by controlling the gap between them~\cite{Zhang18A,Pan24Ultra,Tao24Ultra,Wang24A}.  
Cavity $b$ is built of  $\chi^{(2)}$ nonlinear materials~\cite{Wang21Efficient,Zhang19Broadband,Lu21Efficient}. 
To date, a wide range of nonlinear materials~\cite{Wang21Efficient,Zhang19Broadband,Lu21Efficient}, such as silicon nitride, aluminum nitride, and lithium niobate  have been used to fabricate ultra-high quality factors WGM microcavities with large $\chi^{(2)}$ nonlinearity. 
Both cavities support clockwise (CW) and counterclockwise (CCW) optical modes with resonator frequency $\omega_a$ and $\omega_b$, and decay rate $\kappa_a$ and $\kappa_b$, respectively. 
A two-level atom with transition frequency $\omega_\sigma$ and decay rate $\gamma$ is coupled to cavity $a$ with strength $g$, and is driven by a strong driving field with frequency $\omega_p$ and strength $\Omega_\sigma$. 
Additionally, a continuous laser with frequency $\omega_d$ and amplitude $\epsilon_d$ is applied in the resonator $b$ from port 3, generating a directed squeezing in the CW mode~\cite{Tang22Quantum}. 
Owing to the $\chi^{(2)}$ nonlinear interaction, the mode $b_{cw}$ can be converted to a squeezed mode $b_{cw,s}$, while the mode $b_{ccw}$ is unsqueezed [see Figs.~\ref{Fig1} (c,d)]. 
To detect the nonreciprocal photon bundle emission, a weak probe laser with frequency $\omega_p$ and strength $\Omega_a$ is input in the cavity $a$ from port 1 or 2, corresponding to the left input case or right input case. 

Therefore, in units of $\hbar=1$ and in a frame rotating at frequency $\omega_p$, the Hamiltonian of the system for left input case is given by
\begin{align}
	H_L=\ &\Delta_a a_{cw}^\dag a_{cw} + \Delta_b b_{ccw}^\dag b_{ccw} + \Delta_\sigma \sigma_{+} \sigma_{-} \nonumber\\&+ J(a_{cw}^\dag b_{ccw}+ b_{ccw}^\dag a_{cw}) + g(a_{cw}^\dag \sigma_{-}+a_{cw} \sigma_{+}) \nonumber\\&+ \Omega_a(a_{cw}^\dag + a_{cw}) + \Omega_\sigma(\sigma_{+} + \sigma_{-}),
	\label{Eq1}
\end{align}
where $\Delta_{a(b,\sigma)} = \omega_{a(b,\sigma)}-\omega_p$ denotes the frequency detuning between the cavity $a$ (cavity $b$, atom) and the probe field. 
The operators $a_{cw}$ and $b_{ccw}$ represent the annihilation operators of photons for the CW mode in cavity $a$ and the CCW mode in $b$, respectively.
The $\sigma_{+}=|e\rangle\langle g|$ ($\sigma_{-}=|g\rangle\langle e|$) is the raising (lowering) operator for the two-level atom with the ground state $|g\rangle$ and the excited stated $|e\rangle$.  
The parameter $J$ is the coherent coupling strength between the two resonators, which can be adjusted by adjusting the gap between them.

For the right input case, the effective Hamiltonian reads~\cite{Tang22Quantum}
\begin{align}
	H_R=\ &\Delta_a^d a_{ccw}^\dag a_{ccw} + \Delta_b^d b_{cw}^\dag b_{cw} + \Delta_\sigma^d \sigma_{+} \sigma_{-} \nonumber\\&+ J(a_{ccw}^\dag b_{cw}+ b_{cw}^\dag a_{ccw}) + g(a_{ccw}^\dag \sigma_{-}+a_{ccw} \sigma_{+}) \nonumber\\&+ \frac{\Omega_d}{2}(b_{cw}^{\dagger 2}e^{-i\theta_d}+ b_{cw}^{2}e^{i\theta_d})+ \Omega_\sigma(\sigma_{+} e^{-i\Delta t}+ \sigma_{-}e^{i\Delta t})  \nonumber\\&+\Omega_a(a_{ccw}^\dag e^{-i\Delta t} + a_{ccw}e^{i\Delta t}),
	\label{Eq2}
\end{align}
where $a_{ccw}$ and $b_{cw}$ are the annihilation operators of photons for CCW mode in cavity $a$ and CW mode in $b$, respectively. 
$\Delta_{a(b,\sigma)}^{d}=\omega_{a(b,\sigma)}-\omega_d/2$ represent the frequency detuning  between the cavity $a$ (cavity $b$, atom) and the driving field from port 3.  
$\Delta=\omega_p-\omega_d/2$ is the frequency detuning of the probe and the driving field. 
The amplitude $\Omega_d$ and phase $\theta_d$ are created and determined by the driving field from port 3~\cite{Tang22Quantum}.

\begin{figure*}[t!]
	\includegraphics[width=0.95\linewidth]{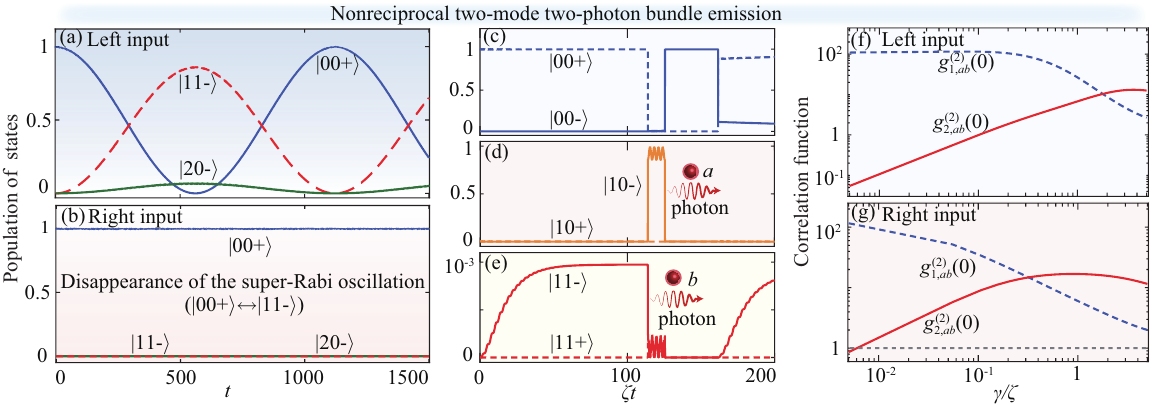}
	\centering
	\caption{Nonreciprocal two-photon bundle emission with the photons emitted from different cavities.
		(a,b) Nonreciprocal super-Rabi oscillation between the states $|00+\rangle$ and $|11-\rangle$. 
		Figures~(a) and (b) depict the populations of the states $|00+\rangle$, $|11-\rangle$, and $|20-\rangle$ as a function of scaled evolution time $\xi t$, when the probe field is input from the left side and the right side, respectively.  
		(c-e) Quantum trajectory of the populations of the states  $|nm\pm\rangle$ ($n,m=0,1$) for the probe field input from left side. 
		(f,g) The cross second-order correlation function $g_{1,ab}^{(2)}(0)$ and the  generalized cross second-order correlation function $g_{2,ab}^{(2)}(0)$ as functions of the atom decay rate $\gamma/\xi$.
		The parameters are selected as $g = \xi, \Omega_a = 0.001\,\xi$, $\kappa_a = \kappa_b = 0.08\xi$, $\gamma = 0.1\kappa_a$, and the other parameters are the same as those in Fig.~\ref{Fig1}.}
	\label{Fig2}
\end{figure*}

Applying the Bogoliubov transformation~\cite{Tang22Quantum},  $b_{cw,s}=\mathrm{cosh}(r)b_{cw}+e^{-i\theta_d}\mathrm{sinh}(r)b_{cw}^\dagger$ with the squeezing parameter  $r=0.25\text{ln}[(1+\beta)/(1-\beta)]$ with $\beta=\Omega_d/\Delta_b^d$, and the frame rotating at frequency $\Delta$, the Eq.~(\ref{Eq2}) reduces to
\begin{align}
	H_R^s=\ &\Delta_a a_{ccw}^\dag a_{ccw} + \Delta_{b,s}^p b_{cw,s}^\dag b_{cw,s} + \Delta_\sigma \sigma_{+} \sigma_{-} \nonumber\\&+ J_s(a_{ccw}^\dag b_{cw,s}+ b_{cw,s}^\dag a_{ccw}) + g(a_{ccw}^\dag \sigma_{-}+a_{ccw} \sigma_{+}) \nonumber\\&+\Omega_a(a_{ccw}^\dag  + a_{ccw}) + \Omega_\sigma(\sigma_{+} + \sigma_{-}).
	\label{Eq3}
\end{align}
In Eq.~(\ref{Eq3}), we have used the rotating-wave approximation and neglected the counter-rotating wave term,  $-Je^{i\theta_d}sinh(r)a_{ccw}b_{cw,s}-Je^{-i\theta_d}sinh(r)a_{ccw}^{\dagger}b_{cw,s}^{\dagger}$. 
This is valid, since the coupling rate much smaller than the natural frequencies of the cavity mode parts. 
Thus, the counter-rotating terms oscillate at much
higher frequencies than the natural dynamics of the system. 
This allows us to treat the rapid oscillations as rapidly
approaching zero, so it is a reasonable approximation to ignore them. $\Delta_{b,s}^{p}=\Delta_{b,s}^{d}-\Delta$ with $\Delta_{b,s}^{d}=\Delta_{b}^{d}(1-\beta)e^{2r}$, and $J_s=J\mathrm{cosh}(r)$. 
A comparison of the Eq.~(\ref{Eq1}) and Eq.~(\ref{Eq3}) reveals that the two Hamiltonian have same form but differ crucially in frequency detuning of cavity $b$ and the coupling strength between the two resonators. 
In addition, the parameter difference is tunable via the squeezing parameter $r$. 
For $r=0 (\beta=0)$, the Eq.~(\ref{Eq3}) can reduces to the Eq.~(\ref{Eq1}). 
The direction-dependent parameter difference can lead to the nonreciprocal photon bundle emission substantially.

We now analyze the eigensystem of the system based on Eq.~(\ref{Eq1}) under the Mollow regime. 
This regime is established when the coupling strengths $J$ and $g$ are much weaker than the atomic driving strength $\Omega_\sigma$. 
Under this condition, the strong pump dresses the atomic levels, thereby forming a Mollow ladder~\cite{Munoz14Emitters} and the weaker couplings can  be treated as a perturbation. 
Within this framework, we demonstrate the nonreciprocal super-Rabi oscillation.
In the absence of the perturbations, we can obtain the eigenstates 
\begin{align}
	|E\rangle_\pm =|n_an_b\pm\rangle,
	\label{Eq4}
\end{align}
where the $n_a$ and $n_b$ are the photon number in the two cavities, respectively, and
\begin{align}
	|\pm\rangle = \ & C_\pm|g\rangle \pm C_\mp |e\rangle, \nonumber\\
	C_\pm = \ & \sqrt{\frac{2\Omega_\sigma^2}{\Delta_\sigma^2 + 4\Omega_\sigma^2\pm \Omega_\sigma \sqrt{\Delta_\sigma^2+4\Omega_\sigma^2}}}.
	\label{Eq5}
\end{align}
The corresponding eigenvalues is
\begin{align}
	E_{n_a,n_b,\pm} = n_a\Delta_a+n_b\Delta_b+\frac{\Delta_\sigma \pm \sqrt{\Delta_\sigma^2 + 4\Omega_\sigma^2}}{2}.
	\label{Eq6}
\end{align}
Eqs.~(\ref{Eq4},\ref{Eq6}) reveal that when the probe field frequency is resonant with the transition from $|00+\rangle$ to $|nm-\rangle$ ($n,m$ represent the number of photons in cavities $a$ and $b$, respectively), the JC interaction and coupling between the two resonators induce the transition between them. 
Meanwhile, transitions to higher-energy levels are suppressed by the frequency detuning.

Thus, when the probe field is input from the left side and frequency is resonant with $|00+\rangle$ to $|11-\rangle$, i.e., $\sqrt{\Delta_\sigma^2+4\Omega_\sigma^2}-\Delta_a-\Delta_b=0$, the resonator transition between $|00+\rangle \leftrightarrow |11-\rangle$ can be induced. 
This is confirmed in Fig.~\ref{Fig2}(a), which shows the populations of the states $|00+\rangle$, $|11-\rangle$, and $|20-\rangle$ as functions of the scaled evolution time $\xi t$, by using the initial state $|00+\rangle$. 
The results clearly show a super-Rabi oscillation between the initial state $|00+\rangle$ and the photon-photon state $|11-\rangle$ for the left input. 
In contrast, when the probe field is input from right side, there is no transition between the two states, as shown in Fig.~\ref{Fig2}(b). 
This demonstrates that nonreciprocal super-Rabi oscillation can be achieved via directional quantum squeezing. 

The physical mechanism underlying the directional super-Rabi oscillation can be understood as follows. 
Since the similar form of the Hamiltonian exists in different driving direction, the anharmonic energy spectrum is responsible for the nonreciprocal super-Rabi oscillation. 
As illustrated in Fig.~\ref{Fig1}(e), a probe field input from the left side resonantly drives the transition of $|00+\rangle \leftrightarrow |11-\rangle$. 
This initiates a super-Rabi oscillation due to the JC interaction and coupling between the two resonators. 
The anharmonicity suppresses transitions to higher energy levels due to detuning. 
However, when the probe field is input from the right side [see Fig.~\ref{Fig1}(f)], the squeezing interaction occurs within the system and alters the effective coupling between the resonators. 
This breaks the frequency-matching condition and consequently inhibits the super-Rabi oscillation.   
This quantum-squeezing-enabled nonreciprocity indicates the generation of nonreciprocal bundled photon pairs, as the super-Rabi oscillation is a prerequisite for multiquanta emission. 

\section{Nonreciprocal bundle emission of photon-photon pairs}\label{section5}
In this section, we demonstrate that nonreciprocal photon bundle emission can be achieved by introducing suitable dissipation into the system. 
In the presence of photon and atomic dissipation, the dynamics of the system are governed by the master equation~\cite{Gardiner00Quantum}
\begin{align}
	\frac{d\rho}{dt}=-i[H,\rho]+\kappa_a\mathcal{L}[a_\zeta]\rho +\kappa_b\mathcal{L}[b_\zeta]\rho+\gamma\mathcal{L}[\sigma_-]\rho,
	\label{Eq7}
\end{align}
where the $H=H_L$ or $H_R^s$ means that the probe field is input into the system from the left or right side. 
For the left input case, we have $a_\zeta= a_{cw}$ and $b_\zeta= b_{ccw}$, while $a_\zeta= a_{ccw}$ and $b_\zeta= b_{cw,s}$ represents the right input case. 
$\rho$ is the density operator of the system. $\mathcal{L}[o]\rho = (2o\rho o^\dagger - \rho o^\dagger o - o^\dagger o \rho)/2$ with $o=a_\zeta$, $b_\zeta$ and $\sigma_-$ is the Lindblad term. 
In the above master equation, thermal effects are neglected. 
Since the resonance frequency of the cavities is on the order of $10^{15}\mathrm{Hz}$, the thermal mean occupation number $n_{a,b}^{th}=[\mathrm{exp}(\hbar \omega_{(a,b)}/k_BT)]^{-1}$ (where $k_B$ is the Boltzmann constant and $T$ is the bath temperature of the optical mode) is very low. 
For instance, at room temperature, the $n^{th}$ is approximately $10^{-14}$.  
Furthermore, for the backward-input case in the system, the continuous pump field in the squeezing configuration inevitably increases the noise in cavity $b$. 
As shown in Ref.~\cite{Lu15Squeezed}, a broadband squeezed-vacuum field, input the system from port 3, can be used to suppress the squeezing-induced noise.  
Consequently, when the probe is input from the right side, cavity $a$ couples to a squeezed mode of cavity $b$ without the squeezing-induced noise~\cite{Tang22Quantum}. 

\begin{figure*}[t!]
	\includegraphics[width=0.95\linewidth]{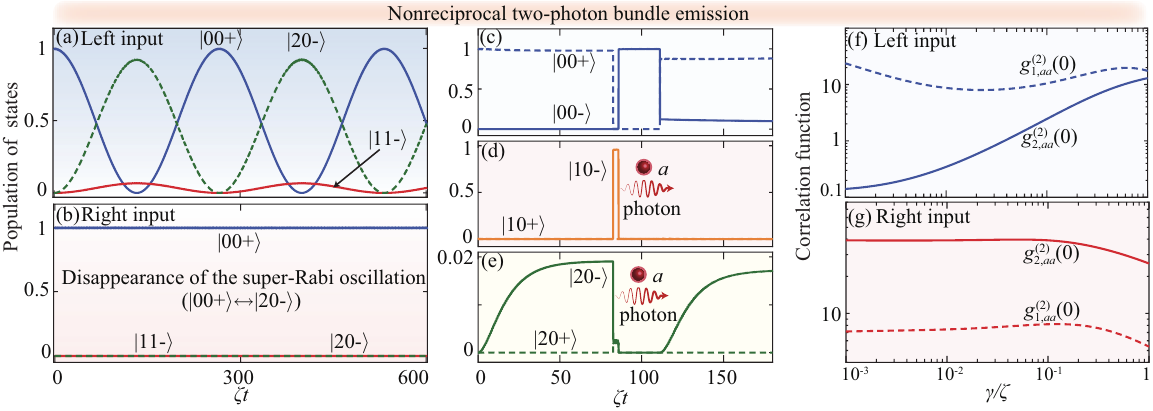}
	\centering
	\caption{Nonreciprocal two-photon bundle emission with photons  emitted from cavity $a$. 
		(a,b) Nonreciprocal super-Rabi oscillations between states $|00+\rangle$ and $|20-\rangle$. 
		Populations of the states $|00+\rangle, |11-\rangle$, and $|20-\rangle$ as functions of the scaled evolution time $\xi t$ for the probe field input from the left side (a) or the right side (b).  
		(c-e) Quantum trajectory of populations of the states $|nm\pm\rangle$ ($n,m=0,1,2$) for the probe field input from left side. 
		Second-order correlation function $g_{1,aa}^{(2)}(0)$ and generalized second-order correlation function $g_{2,aa}^{(2)}(0)$ as functions of the atom decay rate $\gamma/\xi$ for the probe field incident from the (f) left side or the (g) right side. 
		The probe field frequency is resonant with the transition between $|20-\rangle$ and $|00+\rangle$, and the driving parameter from port 3  is set as $\beta = 0.98$. 
		The other parameters are the same as those in Fig.~\ref{Fig2}.}
	\label{Fig4}
\end{figure*}

When the probe field is incident from the left side,  super-Rabi oscillations occur. 
With the introduction of dissipation, photons can be emitted, leading to the emergence of photon bundle emission in this system.  
To illustrate the photon-pair emission process more clearly, Figs.~\ref{Fig2}(c-e) display the quantum trajectories of the populations for the states $|00\pm\rangle$,$|10\pm\rangle$,$|11\pm\rangle$, by using the Monte Carlo simulation.  
The system is initially prepared in the state $|00+\rangle$, with the probe frequency resonant with the transitions $|11-\rangle$ and $|00+\rangle$. 
Under the Jaynes–Cummings interaction and the coupling between the two cavities, the two-photon state $|11-\rangle$ is occupied rapidly, as shown in Fig.~\ref{Fig2}(e). 
Triggered by dissipation, a photon is emitted from cavity $b$, collapsing the wave function to the one-photon state $|10-\rangle$, which is confirmed by the nearly unit population probability of this state, shown in Fig.~\ref{Fig2}(d).
Subsequently, another photon is emitted from cavity $a$ within a short time window, collapsing the system to the zero-photon state $|00-\rangle$ and resulting in the photon-photon bundle emission, as shown in Fig.~\ref{Fig2}(c). 
Given that the coefficients satisfy $|C_+|^2 \gg |C_-|^2$, the superposition states of the atom 
$|\pm\rangle$ can be effectively regarded as the states 
$|e\rangle$ and $|g\rangle$. 
Furthermore, due to spontaneous emission of the atom, a photon at frequency $\omega_\sigma$ is emitted, returning to the initial state of the system and preparing the next photon bundle emission. 
Thus, in each cycle, the system emits two strongly correlated photons and another photon with different frequency which does not affect the bundle emission process. 
However, when the probe field is incident from the right side, the photon bundle emission disappears due to the absence of super-Rabi oscillations.

The statistical properties of photon bundles can be characterized by the following correlation function.   
The statistical properties between photons can be quantified by the second-order correlation function $g_{1,oo'}^{(2)}(\tau)= \langle o^\dagger(0) o'^\dagger(\tau) o'(\tau) o(0) \rangle / [\langle o^\dagger o(0)\rangle \langle o'^\dagger o'(\tau)\rangle]$~\cite{Scully97Quantum}. 
Given a steady state, values of $g_{1,oo'}^{(2)}(0)>1$ and $g_{1,oo'}^{(2)}(0)<1$ correspond to super-Poissonian (bunching) and sub-Poissonian (antibunching) photon statistics, respectively.  
Moreover, correlations among photon-pair bundles themselves can be described using the generalized second-order correlation function~\cite{Munoz14Emitters}
\begin{align}
	g_{2,oo'}^{(2)}({\tau})=\frac{\langle  (oo')^\dagger (0) (oo')^\dagger (\tau) (oo') (\tau) (oo') (0)  \rangle}{\langle (oo')^\dagger  (oo') (0)  \rangle \langle (oo')^\dagger  (oo') (\tau)  \rangle},
	\label{Eq8}
\end{align}
where the $\tau$ denotes the time delay. 
Therefore, given a steady state, $g_{1,oo'}^{(2)}(0)>1$ and $g_{2,oo'}^{(2)}(0)<g_{2,oo'}^{(2)}(\tau)$ [$=g_{2,oo'}^{(2)}(\tau)$,  $>g_{2,oo'}^{(2)}(\tau)$] correspond to antibunching (coherence, bunching) of the photon pairs.

To reveal the nonreciprocal statistical properties of the photon pairs bundle, we plot the cross correlation functions $g_{1,ab}^{(2)}(0)$, $g_{2,ab}^{(2)}(0)$ as functions of the decay rate of the atom $\gamma$ for opposite probe field directions in  Figs.~\ref{Fig2}(f,g). 
In the absent of directional squeezing, the cross correlation functions are reciprocal regardless of the direction of probe light, as the entire system remains symmetric.  
In contrast, the cross correlations exhibit giant nonreciprocity with the directional quantum squeezing.  
As shown in Fig.~\ref{Fig2}(f), when the probe field enters system from the left side, the antibunching of photon pairs can be generated with $g_{1,ab}^{(2)}(0)>1$ and $g_{2,ab}^{(2)}(0)\ll1$. 
However, for the reverse input direction, the bunching of photon pairs occur with $g_{1,ab}^{(2)}{(0)}>1$ and $g_{2,ab}^{(2)}(0)>1$. 
The mechanism underlying the photon pairs bundle emission can be explained as follows. 
For the left input case, super-Rabi oscillations are established. 
Assisted by dissipation, the system can emit two strongly-correlated photons continuously, achieving the two-photon bundle emission. 
In the case of right side input, however, quantum squeezing induces a shift in the eigenenergy levels, suppressing the super-Rabi oscillation. 
Thus, the asymmetric frequency detuning between the probe field and the resonant transition $|00+\rangle \leftrightarrow |11-\rangle$ is responsible for the  observation of strong nonreciprocal photon pairs bundle. 

By optimizing the system parameters further, an additional type of nonreciprocal two-photon bundle emission can also be realized. 
With the probe frequency resonant between states  $|20-\rangle$ and $|00+\rangle$ and the driving parameter from port 3 is set to $\beta=0.98$, we plot the populations of states $|20-\rangle$, $|00+\rangle$ and $|11-\rangle$ as functions of the scaled time $\xi t$ in  Figs.~\ref{Fig4}(a,b). 
It can be seen that a super-Rabi oscillation between the states $|00+\rangle$ and $|20-\rangle$ is clearly established for the left input side, while it is strongly suppressed for the right input case.  
When the two-photon super-Rabi oscillation is sustained, dissipation can trigger the successive emission of two photons from cavity $a$ within a short time window, giving rise to two-photon bundle emission, as shown in Figs.~\ref{Fig4}(c-e). 

The nonreciprocity of two-photon bundle emission can be further confirmed via correlation functions.  
Figures~\ref{Fig4} (f,g) present the correlations $g_{1,aa}^{(2)}(0)$ and $g_{2,ab}^{(2)}(0)$ as functions of the atom decay rate $\gamma$ under opposite propagation directions of the probe field.  
It can be seen that when the probe field is input from the left side, the conditions $g_{1,aa}^{(2)}(0)>1$ and $g_{2,aa}^{(2)}(0)\ll1$, indicate the emission of antibunched two-photon bundles from cavity $a$. 
In contrast, for right input case, both $g_{1,aa}^{(2)}(0)>1$ and $g_{2,aa}^{(2)}(0)>1$ are satisfied, corresponding to bunched photon pairs. 
This nonreciprocal behavior stems from the directional quantum squeezing, which induces a chiral interaction between the two resonators and leads to an asymmetric shift in eigenenergy levels for opposite probe directions.

\section{Conclusions} \label{C}

In summary, we have demonstrated that introducing directional quantum squeezing enables robust nonreciprocal photon bundle emission in a compound system comprising two coupled resonators and a two-level atom. 
Mediated by directional squeezing, an asymmetric photon-hopping interaction between the two resonators arises for opposite propagation directions of the probe field. 
This leads to a corresponding asymmetric eigenenergy levels in opposite probe field direction.  
By analyzing the two-photon state population, we show that directional quantum squeezing enables selective super-Rabi oscillation for the probe field input from one side. 
Furthermore, aided by intrinsic dissipation, the system gives rise to two distinct types of nonreciprocal two-photon bundle emission. 

These findings extend the exploration of directional quantum squeezing induced nonreciprocity into the regime of multiquanta states, building a marriage of nonreciprocal physics and multiquanta emission. 
Photon bundle emission is an effective way to achieve multiquanta states. 
Nonreciprocity provides a unique ability to control the direction of energy flow. 
Thus, nonreciprocal photon bundle emission can achieve more precise control over entangled photon pairs, ensuring that photon pairs are generated in a specific direction and under controlled conditions. 
This strategy employs an all-optical scheme to achieve nonreciprocal multiquanta emission, requiring only single-cavity two-mode matching and eliminating moving parts. 
Furthermore, it utilizes quantum squeezing, an experimentally well-established technique for enhancing light-matter coupling, to ensure seamless compatibility with integrated photonic platforms. 
Thus, our approach can be integrated with established nonreciprocal strategies to advance nonreciprocal engineering.  
This holds the potential for applications in chiral quantum science and technologies.
In a broader context, the proposed scheme may inspire further all-optical investigations into other nonreciprocal quantum phenomena, including nonreciprocal quantum superposition~\cite{Li23Optomechanical,Li23Nonreciprocal}, nonreciprocal soliton~\cite{Li21Nonreciprocal}, and nonreciprocal quantum sensing~\cite{Wang24Quantum}.

\begin{acknowledgments}
We thank Yunlan Zuo and Qian Zhang for helpful discussions and suggestions. H.J. is supported by the National Key R\&D Program of China (No. 2024YFE0102400), the National Natural Science Foundation of China (NSFC, Grant No. 12421005), the Hunan Major Sci-Tech Program (2023ZJ1010), and Quantum Science and Technology-National Science and Technology Major Project (2024ZD0301000). 
L.-M.K. is supported by the NSFC (Grants No. 12247105 and 12421005), Quantum Science and Technology-National Science and Technology Major Project (Grant No. 2024ZD0301000), the Hunan Provincial Major Sci-Tech Program (Grant No. 2023ZJ1010), and the Henan Science and Technology Major Project (Grant No. 241100210400).  
C.L. is supported by the Quantum Science and Technology-National Science and Technology Major Project (Grant No. 2025ZD0300801), and the National Natural Science Foundation of China (Grants Nos. 92476201). 
B.L. is supported by the NSFC (Grant No. 12347136) and the Postdoctoral Fellowship Program (Grade C) of China Postdoctoral Science Foundation (Grant No. GZC20231726).   
T.-X.L. is supported by the NSFC (Grant Nos. 12205054, 12565001) and the Natural Science Foundation of Jiangxi Province (No. 20252BAC200163). 
J.-X.L. is supported by the Henan Province Postdoctoral Research Funding Project (Grant No. HN2025011)
\end{acknowledgments}

\bibliographystyle{unsrt}

\end{document}